# Genesis of Dark Energy: Dark Energy as a Consequence of Cosmological Nuclear Energy

R.C.Gupta
Professor
Institute of Engineering & Technology (I.E.T.)
Lucknow-226021, India
rcg_iet@hotmail.com

**Abstract**

Recent observations on Type-Ia supernovae and low density ( $\Omega_m$=0.3) measurement of matter including dark matter suggest that the present day universe consists mainly of repulsive-gravity type 'exotic matter' with negative-pressure often referred as 'dark energy'($\Omega_x$=0.7). But the mystery is about the nature of dark energy and its puzzling questions such as why, how, where & when about the dark energy are intriguing. In the present paper the author attempts to answer these questions while making an effort to reveal the genesis of dark energy, and suggests that the cosmological nuclear binding energy liberated during primordial nucleo-synthesis remains trapped for long time and then is released free which manifests itself as dark energy in the universe. It is also explained why for dark energy the parameter w = -2/3. Noting that w=+1for stiff matter and w=+1/3 for radiation; w = - 2/3 is for dark energy, because '-1' is due to 'deficiency of stiff-nuclear-matter' and that this binding energy is ultimately released as 'radiation' contributing '+1/3', making w = -1 + 1/3 = -2/3. This thus almost solves the dark-energy mystery of negative-pressure & repulsive-gravity. It is concluded that dark energy is a consequence of released-free nuclear energy of cosmos. The proposed theory makes several estimates / predictions, which agree reasonably well with the astrophysical constraints & observations.

## 1. Introduction

It is an irony of nature and puzzling that most abundant form of matter-energy in the universe is also most mysterious. Supernovas at relatively high red shift are found fainter than that predicted for an earlier-thought slowing-expansion and indicate that expansion of universe is actually speeding up [1-3]. Recent studies [3,4] firmly establish that universe is now undergoing an acceleration; with repulsive gravity of some strange energy-form i.e., dark-energy at work. Dark energy, a 'mysterious substance' whose pressure is 'negative' and accounts for 70% of total matter-energy budget of the universe, but has no clear explanation. Understanding its origin & nature is one of the greatest problems of present time. In the present paper; a modest approach has been made to explain the genesis of this dark energy, suggesting that dark-energy is a result of released-free cosmological nuclear-energy in the universe.

## 2. Cosmo-Dynamics

The well known Friedman Equations and the analysis for cosmo-dynamics are briefly summarized as follows for clarity & completeness and for its subsequent uses:

$$a'^2/a^2 + k/a^2 = 1/3\ 8\pi G\ \rho \qquad (1)$$

$$2a''/a + a'^2/a^2 + k/a^2 = -\ 8\pi G p \qquad (2)$$

From Eqs.1 & 2, the following equations are written as,

$$2\ a''/a = -1/3\ 8\pi G\ (\rho + 3p) \qquad (3)$$

$$\rho' = -\ 3(\rho + p)\ a'/a \qquad (4)$$

where scale-factor or universe-size at a particular time is a(t) or simply as a . Gravitational-constant is G and ρ is the total density $\rho = \rho_r + \rho_m + \rho_x$ where subscripts r is for radiation, m is for matter (including dark matter) & x is for dark energy. The curvature factor is k and depending on k being 1, 0, -1 the universe is closed, flat or open. First or second differential of a variable, usually indicated with single or double dot over it, are indicated here with single or double dash on it.

The Hubble's law for universe-expansion is as follows (Eq.5); and from Eqs 1 & 5, Eq.6 can be written as follows, H being the Hubble's constant :

$$a^{/} = H.\ a \tag{5}$$

$$H^2 + k/a^2 = 1/3\ 8\pi G\ \rho \tag{6}$$

Critical density $\rho_c$ is defined as that density for which the universe is flat (k = 0), thus from Eq.6,

$$\rho_c = 3H^2/\ (8\pi G) \tag{7}$$

The deceleration parameter q is defined as,

$$q = -\ a^{//}.a\ /a^{/\ 2} = -a^{//}\ /\ (aH^2) \tag{8}$$

The equation of state parameter w is defined as,

$$p = w.\ \rho \tag{9}$$

From Eqs. 4 & 9,

$$\rho\ \sim 1/a^{\ 3(1+w)} \tag{10}$$

From Eqs 1 & 10 for flat (k = 0) universe,

$$a \sim t^{\ 2/3(1+w)} \tag{11}$$

Usually a parameter called cosmological red-shift (Z) is often referred as a measure of the era (time t) of the universe. For example; $Z_o$ corresponds to the present time $t_o$ with present universe size (scale) $a_o$ , and Z corresponds to some time t in the past with a smaller universe size a. In fact the cosmological red-shift for spectral-line (wavelength λ) is defined as $Z = d\lambda/\lambda = (a_o - a)/a = a_o/a - 1$ : thus Z also gives a measure of universe size at a particular time, higher Z means smaller universe size in the past and is expressed as,

$$a_o/a = (1 + Z) \tag{12}$$

**Particular Cases**

(i) For w = +1 stiff matter, $\rho \sim 1/a^6$ and $a \sim t^{\ 1/3}$ (13.a & 13.b)

(ii) For w = +1/3 radiation dominated, $\rho \sim 1/a^4$ and $a \sim t^{\ 1/2}$ (14.a & 14.b)

(iii) For w = 0 matter dominated (dust), $\rho \sim 1/a^3$ and $a \sim t^{\ 2/3}$ (15.a & 15.b)

(iv) For w = -2/3 dark energy (quintessence) $\rho \sim 1/a$ and $a \sim t^{\ 2}$ (16.a & 15.b)

(v) For w = -1 vacuum energy ρ = constant (like cosmological constant) (17)

These variations $\rho$ versus a and a versus t of the particular cases are easily found from Eqs. 10 & 11, and are also shown in Fig. 1.a & b.

## 3. Nuclear Genesis of Dark Energy

### 3.1 Negative pressure due to free-release of cosmological nuclear energy

It is suggested that dark energy is a consequence of nuclear energy. The fusion nuclear binding energy begins to be liberated (but trapped) during primordial nucleo-synthesis, and much later (Z = 80) after decoupling, the nuclear binding energy is released free and appears effectively as dark energy. As shown (Fig.2) in subsequent sections of the paper; that this dark energy is produced earlier but is released free later, quantity-wise ($\Omega_x = 0.0003$) insignificant; remains low during galaxy formation (say, at Z = 3, $\Omega_x = 0.1$), becomes 50% at transition (Z = 0.5, $\Omega_x = 0.5$) & emerges as dominant at present time (Z=0, $\Omega_x = 0.7$) , satisfying all the astrophysical constraints [5].

The equation of state parameter w for dark energy (quintessence) is taken as w = -2/3. The reason for w = -2/3 = -1 + 1/3 is explained as follows. Noting that w = +1 for *stiff* matter & w = 1/3 for radiation; '-1' is due to 'deficiency of *stiff*(nuclear)matter' which is manifested as mass-defect (binding-energy) and '1/3' is because this released binding-energy though initially-trapped but is finally released as 'radiation'. Thus considering that the binding-energy due to mass-defect(deficiency) from stiff-nucleus finally comes out as radiation; the equation of state parameter w = -1 + 1/3 = -2/3. The puzzle of negative pressure (or repulsive gravity) is thus almost solved; the negative pressure (or dark energy) is produced due to released-free cosmological nuclear energy !

### 3.2 Variation of matter-density and dark-energy-density much after decoupling era between Z = 80 to Z=0

Stiff matter density decays very fast and vanishes very soon. Radiation density also decays fast and vanishes soon much after decoupling era. Although, as explained in section 3.3 & in Fig.2, the dark-energy is born much earlier in the past at about $Z = 10^{10}$ but remains trapped dormant and decays in constrained way along with the matter and is released free only after Z = 80 to be effective & dominant. So, between Z =80 to Z = 0 variations of radiation-density, matter-density and dark-energy-density (Eqs 14, 15 & 16) are considered as follows (Eqs.18, 19 & 20). The symbol $\rho$ is for density and $\Omega$ for density ratio $\rho/\rho_c$ ; the subscript m, r , x are for matter, radiation, dark energy and the subscript o indicates the values at present time (Z =0). It can be said that for flat universe ($\rho=\rho_c$),

$$\rho = \rho_c = \rho_r + \rho_m + \rho_x \qquad (17.a)$$

$$\Omega_c = \rho/\rho_c , \quad \Omega_r = \rho_r/\rho_c , \quad \Omega_m = \rho_m/\rho_c , \quad \Omega_x = \rho_x/\rho_c \qquad (17.b)$$

$$\Omega_c = 1 = \Omega_r + \Omega_m + \Omega_x \qquad (17.c)$$

where radiation-density, matter-density and dark-energy-density varies (from Eqs.14, 15 & 16) as follows, taking the present values of densities $\rho_{mo} \sim 10^{-28}$ gram/cm$^3$, $\rho_{xo} \sim 2x10^{-28}$ gram/cm$^3$, $\rho_{ro} \sim 10^{-31}$ gram/cm$^3$ or $\Omega_{mo} \sim 0.3$ & $\Omega_{xo} \sim 0.7$ and negligibly small $\Omega_{ro} \sim 0.0003$ ;

$$\rho_r = \rho_{ro}\, a_o^4/a^4 = 0.0003\, \rho_c\, (1 + Z)^4 \qquad (18)$$

$$\rho_m = \rho_{mo}\, a_o^3/a^3 = 0.3\, \rho_c\, (1 + Z)^3 \qquad (19)$$

$$\rho_x = \rho_{xo}\, a_o/a = 0.7\, \rho_c\, (1 + Z) \qquad (20)$$

## 3.3 **Biography of Dark Energy**

Biography of mysterious dark-energy is even more mysterious. Nevertheless a brief biographical description of dark energy is presented here somewhat idiomatically/metaphorically for interesting-clarity.

Dark energy in fact is the released nuclear binding energy of cosmos. Dark energy is born during primordial nucleo-synthesis epoch ($Z \sim 10^{10}$) in the early universe out of 'mother' matter present there, when while the nucleo-synthesis (of say, Helium) liberates the binding energy as the 'child' but the child is trapped in or around mother's lap and remains almost dormant for long period and moves parallel to mother's foot-steps. Approximate weight of this dormant dark-energy (child) is roughly estimated as 1% of binding energy, of 25% of primordial helium-synthesis, from 13% of baryonic-matter ($\Omega_b$=0.04) out of total mass of (mother) matter ($\Omega_m$=0.3); this comes out to be $\rho_x/\rho_m$ = 0.000325 and since this era is radiation dominated era $\rho_r$ being extremely high gives $\Omega_x$ very close to zero. This 'child' i.e., dark-energy (nuclear-binding-energy liberated) though has come out of mother's womb(nucleus) remains trapped / dormant for quite long time atleast upto decoupling era Z =1000 and even beyond when positive radiation ($\rho_r$) pressure is much more dominant than the negative pressure of dark energy radiation ($\rho_x$), the free-release of this dark energy is thus prohibited till then. While the universe is expanding, the dark energy (the child) however decays parallel to matter (mother) foot-steps at the same rate ( $\rho \sim 1/a^3$) from Z =$10^{10}$ to Z = 1000 and even beyond Z < 100 upto Z =80 (Fig.2). At Z =80, though dark energy $\Omega_x$ still being less than $\Omega_r$ (which opposes liberation of the trapped dark energy) but $\Omega_x$ is sufficient enough to fight with $\Omega_r$, the dark energy is ultimately released. Thus at Z =80, the dark energy (child) is released free from mother's constrained-protection. Then onwards (Z<80) the matter density continues to decay at the same rate ($1/a^3$), but the dark energy (with w = - 2/3 = -1 + 1/3, as explained earlier in section 3.1) density decays at much slower rate (1/a) till today (Z=0). Between Z=80 to Z=0, the dark energy density curve ($\rho_x \sim 1/a$) crosses radiation density curve ($\rho_r \sim 1/a^4$) at Z=12 and crosses matter density curve ($\rho_m \sim 1/a^3$) at Z=0.5. Briefly as summary (Fig.2): ' Dark energy is born as a result of liberated binding energy during primordial nucleo-synthesis at $Z \sim 10^{10}$ as a very very small fraction ($\rho_x/\rho_m \sim 0.0003$, $\Omega_x \sim 0$) of total matter-mass, initially (stage-1) it remains trapped/dormant and decays at fast rate ($1/a^3$) alongwith matter till Z=80 ($\rho_x/\rho_m \sim$ 0.0003, $\Omega_x \sim 0.0003$) and finally it is released free (stage-2) & decays slowly (1/a) between Z=80 to Z=0 and becomes more effective & even dominant today ($\rho_x/\rho_m \sim 2$, $\Omega_x \sim 0.7$ )'. In stage-1: Ωx is almost equal to zero; whereas in stage-2: Ωx increases as Z decreases, $\Omega_x$=0.0003 at Z=80, $\Omega_x$=0.01 at Z=12, $\Omega_x$=0.1 at Z=3, $\Omega_x$=0.5 at Z=0.5 and $\Omega_x$=0.7 at Z=0. The '*mantra*' of dark energy hidden in nucleus comes out free and speaks loud as ' -1 + 1/3 = - 2/3 ' .

## 4. **Estimation of Z and $\Omega_x$**

(i) For present time; $Z = Z_o = 0$ , $\Omega_x = 0.7$ , $\Omega_m = 0.3$ (Present known values).

(ii) For matter/dark-energy transition ($\rho_m = \rho_x$ ); Z can be estimated from Eqs. 19 & 20 as $Z_T$ = 0.5 .

(iii) For galaxy formation (say at $Z = Z_g = 3$); $\Omega_x$ can be estimated from Eqs. 17-20 as $\Omega_{xg} = 0.1$ .

(iv) For radiation/dark-energy transition ($\rho_r = \rho_x$); Z can be estimated from Eqs. 18 & 20 as $Z = Z_t$ = 12.

(v) At $Z = Z_f = 80$ when dark-energy is released 'free', $\Omega_x$ can be estimated (in stage-2) from Eqs.17-20 as $\Omega_{xf} = 0.0001$. The time Z=80 is the joining point for stage-1 & stage-2 (Fig.2). In stage-1 $\rho_x/_m \sim 0.0003$ remains constant, and for stage-2 $\rho_x/\rho_m \sim 2/[(1+Z)^2+2\}$; equating these gives Z ~ 80 as meeting point for the two stages.

(vi) At decoupling epoch ($\rho_m = \rho_r$) $Z = Z_{dec}$=1000 , $\Omega_x$ will be nearly half of initial value of $\rho_x/\rho_m$ in stage-1 (Fig.2), i.e., Ωx=0.00015.

(vii) At $Z=Z_n=10^{10}$ during nucleo-synthesis era birth of dark energy takes place. The fusion nuclear binding energy (= 1%) begins to be liberated during primordial Helium synthesis (25%) from 13% (0.13 = 0.04/0.3) of baryonic-matter ($\Omega_{bo}$ = 0.04) out of total matter-mass ($\Omega_{mo}$ = 0.3). Considering the liberated nuclear-binding-energy as the dark-energy, the dark-energy-density is roughly estimated (1% of 25% of 13%) as 0.000325 of the total matter-density. However, since radiation density at that high value of Z (at nucleo-synthesis time $Z=Z_n \sim 10^{10}$) would be much higher than that of matter, the estimated value for $\Omega_{xn}$ would be close-to-zero.

The liberated nuclear-energy as dark-energy, however, is tracked (as mentioned in section 3.3 and in Fig.2) in two stages. In trapped stage-1 (between $Z = 10^{10}$ to Z = 80) the dark-energy is trapped & dormant and varies ($1/a^3$) in a constrained way along with its parent matter; whereas in free stage-2 (between Z =80 to Z=0) the dark-energy is released free & becomes effective/dominant and varies (1/a) at slower rate during universe expansion. At $Z = 10^{10}$ $\Omega_x \sim 0$, at Z =80 $\Omega_x$ =0.0003, at Z=12 $\Omega_x$ =0.01, at Z = 3, $\Omega_x$ = 0.1, at Z = 0.5 $\Omega_x$ = 0.5 and at Z = 0 $\Omega_x$ =0.7 . It may be noted that these estimates are quite reasonable and satisfy very well all the necessary astrophysical constraints [5].

## 5. Estimating deceleration-parameter q

The space dimension for universe with (i) matter (ii) dark-energy and (iii) both these combined are written as follows:

(i) For matter (w = 0) universe

$$a_m = a = K_m t^{2/3} \qquad \text{where } K_m = a_o/t_o^{2/3} \qquad (21.a \& b)$$

(ii) For dark-energy (w = - 2/3) universe

$$a_x = a = K_x t^2 \qquad \text{where } K_x = a_o/t_o^2 \qquad (22.a \& b)$$

(iii) For the universe combined with matter ($\Omega_m$) plus dark-energy ($\Omega_x$)

$$a = \Omega_m.a_m + \Omega_x.a_x = \Omega_m.K_m t^{2/3} + \Omega_x.K_x t^2 \qquad (23)$$

The deceleration-parameter $q = - a^{//}.a/a^{/\,2}$ can be estimated at present time $t_o$ ($\Omega_m$=0.3, $\Omega_x$=0.7) using Eqs. 23 and 21.b & 22.b and is found as $q_o$ = - 0.52 which indicates that the expansion of universe is not decelerating but accelerating. The current known value [3] is, however, $q_o$ = - 0.67 .

In fact, there could be following two approaches/methods/formulae to find the deceleration-parameter $q = -a^{//}.a/a^{/\,2}$ which are described as follows:

1. Conventional Method

The 'deceleration parameter' for 'only matter' universe can easily be evaluated as $q_m$ = ½ and that for 'only dark-energy' universe as $q_x = ½ + 3/2\ w$ . So simple conventional method usually followed assumes that the actual 'deceleration parameter q for the matter plus dark-energy universe' is 'weighted sum of $q_m$ and $q_x$'. Thus,

$$q_o = ½\ \Omega_{mo} + (½ + 3/2\ w)\ \Omega_{xo} \qquad (24)$$

which gives for w = -2/3, $\Omega_{mo}$ =0.3, $\Omega_{xo}$ = 0.7 ; $q_o$ = - 0.20 a value far different from actual known [3] value of $q_o$ = - 0.67 and thus necessitates modifications in evaluation as follows.

2. Modified Method

The author suggests a new or modified method to evaluate the deceleration-parameter. This modified method assumes that actual 'space scale factor (a) for matter plus dark-energy universe' is 'weighted sum of $a_m$ and $a_x$'. Thus the equation for total actual space-size (a) can be written (similar to Eq.23) as,

$$a = \Omega_m . a_m + \Omega_x . a_x = \Omega_m . K_m t^{2/3} + \Omega_x . K_x t^{2/3(1+w)} \qquad (25)$$

The deceleration parameter $q = - a^{//}.a / a^{/2}$ is evaluated using Eqs.25 and 21.b & 22.b in it at present time $t_o$ as,

$$q_o = - [2/9\ \Omega_{mo} - 2/9\ (1+3w)/(1+w)^2\ \Omega_{xo}] / [2/3\ \Omega_{mo} + 2/3\ /(1+w)\ \Omega_{xo}]^2 \qquad (26)$$

which gives for w = - 2/3, $\Omega_{mo}$ = 0.3, $\Omega_{xo}$ = 0.7 ; $q_o$ = - 0.52 (as found before using Eq.23) which is reasonably near to the known [3] observed value of $q_o$ = - 0.67.

Modified-method (Eq.26) seems to be more correct than the conventional-method (Eq.24) since the expansion (acceleration) of universe due to dark-energy ($a \sim t^2$) is much faster than the expansion (deceleration) of universe due to matter ($a \sim t^{2/3}$). This fact is taken into account in the modified-method (in Eqs. 23, 25 ,26; the second terms more dominant). So, the modified method predicts more nearer value of $q_o$ = - 0.52, instead of - 0.20 from conventional method, to the known observed value of - 0.67 .

## 6. **Discussions**

The exact time of birth Z=$10^{10}$ or $10^9$ can not be very precisely specified. It is suggested by the author that nuclear binding energy manifests itself as dark energy, but the beginning of nuclear energy and its completion is not exactly defined/known. There is two-stage tracking (decay) of dark energy (Fig.2). Initially in stage-1 the decay is constrained & fast during which the dark energy remains almost trapped / dormant for long time. Finally in stage-2 the dark energy is released free at Z =80 and decays slow and thus becomes more effective & dominant in due course. The free released nuclear binding energy plays its role as dark energy (with w = - 2/3) creating negative-pressure or repulsive-gravity. Dark energy begins with close-to-zero value of $\Omega_x$ , remains low till galaxy formation (Z=3, $\Omega_x$=0.1), becomes 50% at transition ($\Omega_x$=0.5, Z=0.5) and then after leads to accelerated universe becomes dominant at present time (Z=0, $\Omega_x$=0.7). All these are well in accordance with the astrophysical constraints [5]. The dark-energy variations, in stage-2 ($\rho_x \sim 1/a$) is governed by final/present condition (at Z =0) and in stage-1 ($\rho_x \sim 1/a^3$) is governed by initial/past condition (at Z =$10^{10}$). The meeting point for the two stages (dotted & firm curves in Fig.2) is found at Z = 80. There could be some error in it (Z=80) due to approximations / estimations. Though meeting point is a point of interest in view that at this time the dark-energy is finally released-free, however, any future change in its estimate (other than Z=80) will hardly have any effect on the other values (estimated in section 4) or on the theme & philosophy of the model which will remain almost unchanged.

For estimating the deceleration parameter, the author suggests a more realistic modified-method (Eq.26) which gives an estimate of $q_o$ = - 0.52 against the observed known value of $q_o$ = - 0.67. The agreement is not bad. The discrepancy in the value, however, may be due to following reason: the dark energy as a result of liberation of nuclear binding energy not only takes place during primordial nucleo-synthesis (Z = $10^{10}$ or $10^9$) era but also due to fusion reaction taking place within stars after or around galaxy formation (Z =4 to 0). It may be noted that only 'primordial nucleo-synthesis', which indeed is more significant, has been taken into consideration; but the 'star nucleo-synthesis' will also have some effect on acceleration of universe i.e., on the deceleration parameter in the right direction. Also, this means that nucleo-synthesis (thus liberation of dark energy) is more or less a continuous process.

Though there are many suspects (candidates) such as [6,7] cosmological-constant, vacuum-energy, scalar-field, brane-world etc. as reported in vast-literature for the dark-energy; the proposed model in this paper at least presents a fresh new-candidate (cosmological nuclear-energy) as possible suspect for the dark-energy. All avenues for possible truth must be kept open.

## 7. **Conclusions**

The author proposes a novel model for genesis of dark energy, indicating its origin in nucleus and suggesting that it is the released-free nuclear binding energy of cosmos which manifests itself as dark-energy causing negative-pressure & repulsive-gravity in the universe. It explains why for dark energy w = -2/3. Also, the model describes the biography of the dark energy and tells that $\Omega_x$ begins with a close-to-zero value, remains low as necessary till galaxy formation say at Z=3 $\Omega_x$=0.1, transition at Z=0.5 $\Omega_x$=0.5 and becomes dominant at present time Z=0 Ωx=0.7; satisfying very well all the astrophysical constraints. The author also proposes a modified-method to estimate deceleration parameter and finds that with the proposed model $q_o$ = - 0.53 in reasonable agreement with known observed value. It can be metaphorically said that 'the dark energy is fossil of nuclear reaction that has taken place in the early universe and that the fossil has recently been noticed when the matter-cover over it has diminished in due course'.

## **Acknowledgements**

The author is obliged & wishes to thank Dr. V.B. Johri, Professor Emeritus Lucknow University and Dr. M.S.Kalara, Professor IIT Kanpur for useful advice & suggestions. The author is also thankful to Dr.V.P. Gautam, Dr. Sanjay Mishra, Dr. Balak Das, RK Gautam, Anil, Sanjay, Ruchi, Sanjiv, Chavvi, Veena and Shefali for their help & cooperation. Also, thanks to IET, UPTU, Govt., AICTE and World-Bank for their support for providing facilities and assistance.

## **References**

1. S. Perlmutter et al., Nature, 391, 51, 1998; Astrophysics J., 517, 565, 1999; A.G. Riess et al., Astron. J, 116, 1009, 1998.
2. M.S. Turner, Nucl. Phys. B (proc. Suppl.), 72, 69, 1999.
3. Wendy L Freedman and Michael S. Turner, arxiv:astro-ph/0308418, Aug. 2003.
4. Y. Wang and M. Tegmark, arxiv:astro-ph/0403292, March 2004.
5. V.B. Johri, Physical Review D, vol 63, 103504, 2001.
6. J.P.Ostriker and P.I. Steinhardt, Quintessentia Universe, Scientific American, Jan. 2001.
7. Robert R Caldwell, Physics Web – Physics World – Dark energy, May 2004.

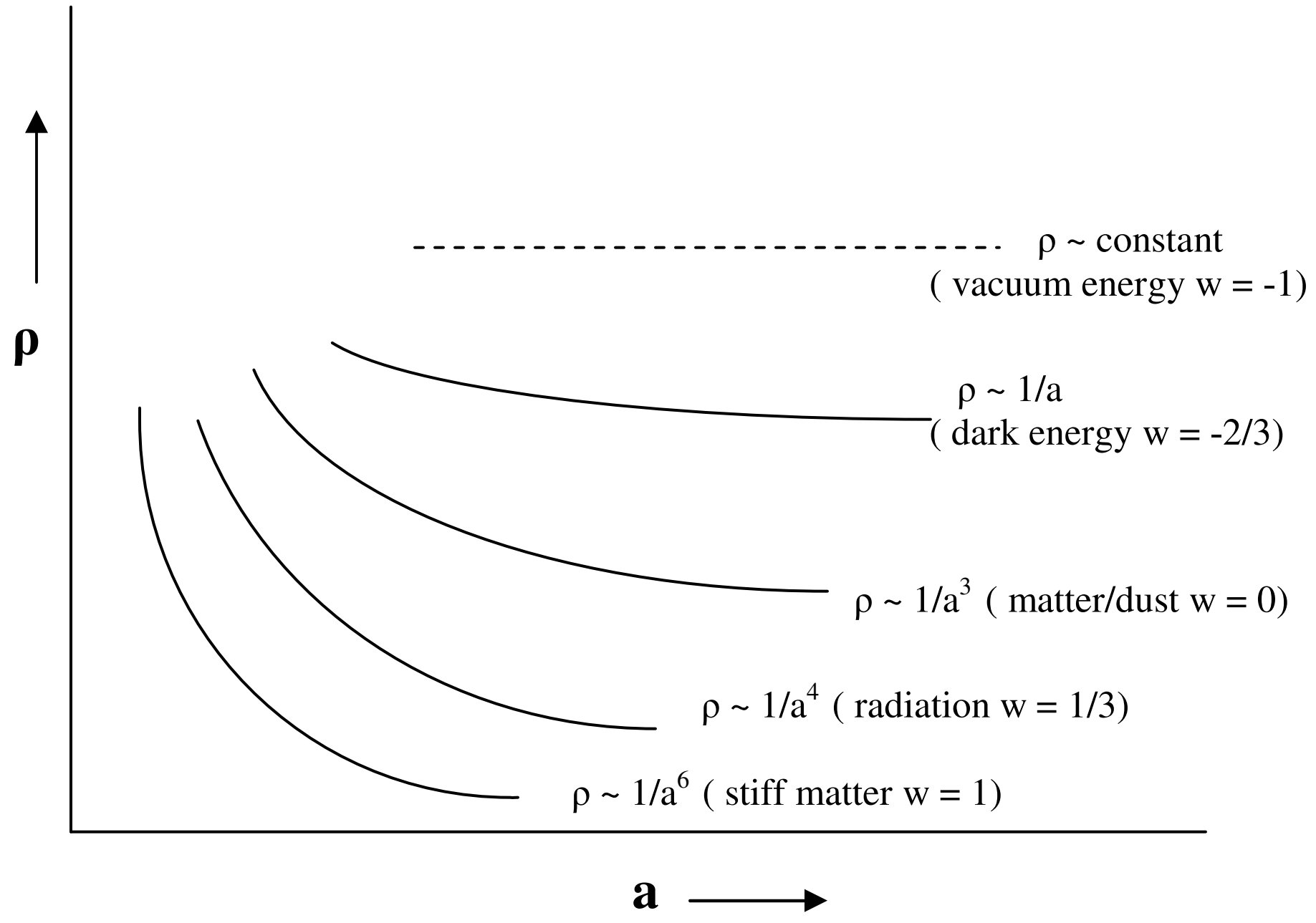

**Fig 1.a** Density **ρ** decay with space size **a**

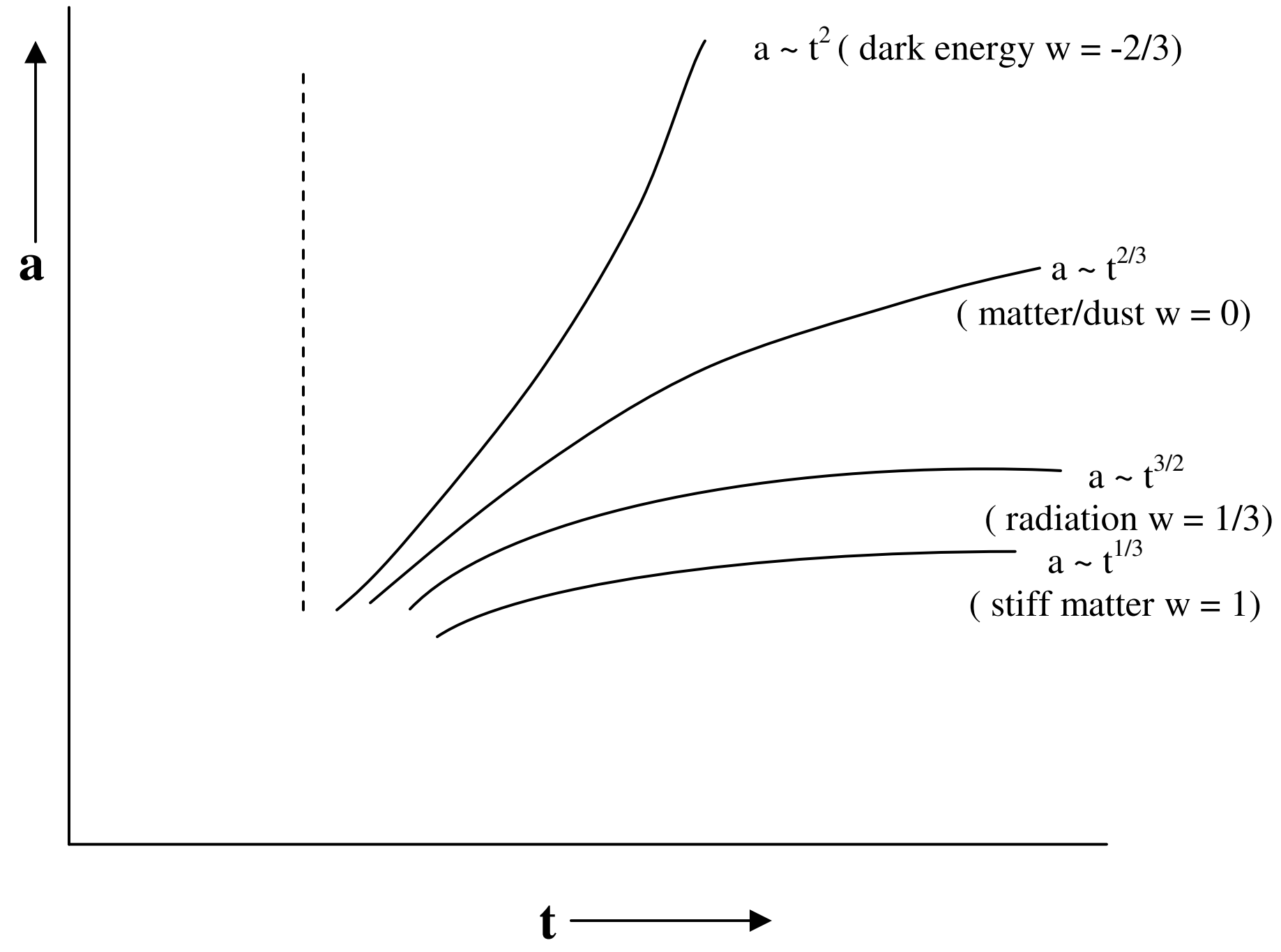

**Fig 1.b** Space size **a** expansion with time **t**

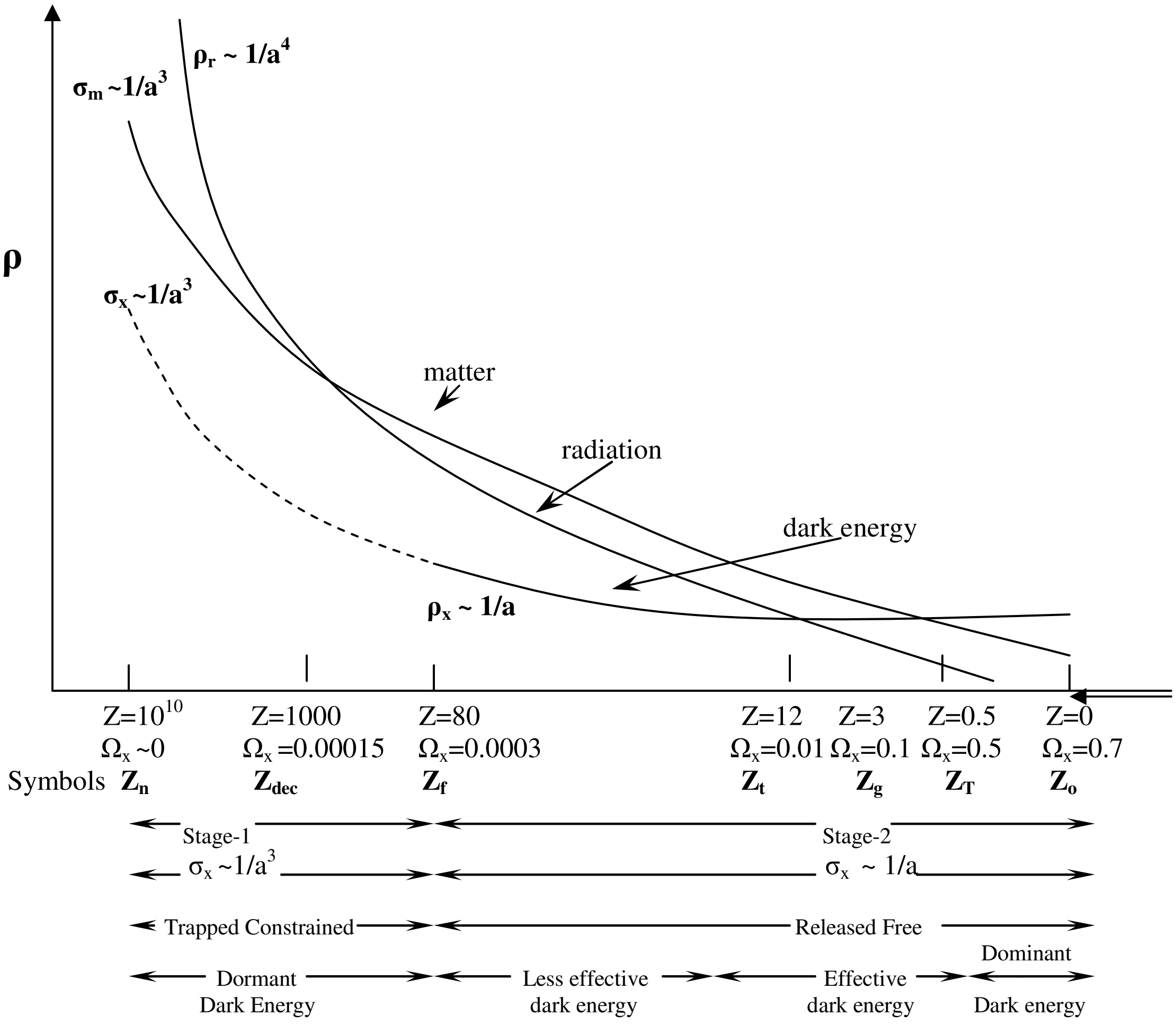

| | |
|---|---|
| $Z_n$ | : Birth of Dormant dark energy (binding energy) at nucleo-synthesis. |
| $Z_{dec}$ | : Decoupling era at $\rho_m = \rho_r$ |
| $Z_f$ | : Dark energy freed and released |
| $Z_t$ | : Transition of dark energy with radiation at $\rho_x = \rho_r$ |
| $Z_g$ | : Galaxy formation ($4 > Z > 2$) say, at Z=3 |
| $Z_T$ | : Transition of dark energy with matter at $\rho_x = \rho_m$ |
| $Z_o$ | : Present time |

**Fig 2** Genesis of Dark Energy $\boldsymbol{\rho_x}$ and its two stage (Dormant and Free) Variations / tracking, along with variations of $\boldsymbol{\rho_r}$ and $\boldsymbol{\rho_m}$